\documentclass[12pt]{iopart}

\usepackage{amssymb}
\usepackage{amsbsy}

\newcommand{\bfm}{\mathbf}
\newcommand{\be}{\begin{equation}} \newcommand{\ee}{\end{equation}}
\newcommand{\bea}{\begin{eqnarray}} \newcommand{\eea}{\end{eqnarray}}
\newcommand{\el}{\nonumber \\}
\newcommand{\re}[1]{(\ref{#1})}
\newcommand{\pat}{\partial}
\newcommand{\adot}{\dot{a}}

\newcommand{\Phidot}{\dot{\Phi}}
\newcommand{\Phiddot}{\ddot{\Phi}}

\newcommand{\Hdot}{\dot{H}}
\newcommand{\bx}{\bi{x}}
\newcommand{\by}{\bi{y}}
\newcommand{\bk}{\bi{k}}
\newcommand{\brt}[1]{[#1]}

\newcommand{\PRD}[1]{{\it Phys. Rev.} {\bf D#1}}
\renewcommand{\PRL}[1]{Phys. Rev. Lett. {\bf #1}}
\newcommand{\NPB}[1]{{\it Nucl. Phys.} {\bf B#1}}
\newcommand{\PLB}[1]{{\it Phys. Lett.} {\bf B#1}}
\newcommand{\MNRAS}[1]{{\it Mon. Not. Roy. Astron. Soc.} {\bf #1}}
\newcommand{\APJ}[1]{{\it Astrophys. J.} {\bf #1}}

\renewcommand{\CQG}[1]{Class. Quant. Grav. {\bf #1}}
\newcommand{\GRG}[1]{{\it Gen. Rel. Grav.} {\bf #1}}

\begin{document}
%\baselineskip16pt

%\begin{titlepage}

\title{Dark energy from backreaction}

\author{Syksy R\"{a}s\"{a}nen}

\address{Theoretical Physics, University of Oxford,
1 Keble Road, Oxford, OX1 3NP, UK}

\ead{syksy.rasanen@iki.fi}

\begin{abstract}

\noindent We consider the effect of inhomogeneities
on the expansion of the Einstein-de Sitter universe.
We find that the backreaction of linear
scalar metric perturbations results in apparent dark
energy with a mixture of equations of state between
0 and --4/3. We discuss the possibility that backreaction
could account for present-day acceleration.

\end{abstract}


%\end{titlepage}

\setcounter{secnumdepth}{3}

\section{Introduction}

\paragraph{The concordance model.}

Perhaps the most surprising observation in recent cosmology
is that the expansion of the universe seems to be accelerating.
This all the more puzzling since the acceleration has apparently
started in the recent past, at a redshift of probably less
than one. These conclusions are based on data from
the cosmic microwave background, large scale structure and supernovae
\cite{SCP, SST, Peebles:2002, Padmanabhan:2002b, Spergel:2003, Tegmark:2003b}.
%[1-6].
Though the only significant direct evidence for acceleration comes
from the supernova observations, data from different sources
seem to fit together. There is also some evidence for acceleration
from correlation of the CMB with large scale structure
\cite{Boughn:2003, Fosalba, Scranton:2003}.
%[7-9].

The preferred framework for interpreting the observations is the 
`concordance model' in which the universe is spatially flat,
and cold dark matter and baryons contribute
$\Omega_{\textrm{m}}\approx1/3$ to the energy density and vacuum energy
contributes the rest, $\Omega_{\Lambda}\approx2/3$.

Most alternatives to the `concordance model' 
replace the cosmological constant with some more
complicated component with negative pressure, called `dark energy'.
Indeed, the relation $\Omega_{\textrm{de}}\approx 2\,\Omega_{\textrm{cdm}}$ suggests
a connection between dark energy and dark matter, and motivates the
construction of models of unified, or coupled, dark matter and dark energy
\cite{antifriction, Schwarz:2002, Kamenshchik:2001, cosmon, Padmanabhan:2002a, Bassett:2002, Comelli:2003}.
%[10-16].

However, the conclusion that some component with negative pressure
is needed at all is prior-dependent
\cite{Elgaroy:2003, Blanchard:2003}: the possibility that a
model with no exotic ingredients could fit the data
as well is not excluded. As a notable example, a model
with $\Omega_{\textrm{m}}=0.88$ and neutrino energy density
$\Omega_{\nu}=0.12$ can fit the CMB and large scale structure
data even better than the `concordance model', though it
does not fit the data from supernovae \cite{Blanchard:2003}. The
neutrino component is needed, since in spatially flat models
with $\Omega_{\textrm{m}}=1$ the density perturbation amplitude $\sigma_8$
seems to be generically too high.

There are two main motivations for looking for alternatives.
First, the `concordance model' does not fit all of the data
very well. In particular, the prediction for the amplitudes of
the quadrupole and octopole of the CMB is too high\footnote{The low multipoles are susceptible to contamination from the Galaxy, but this is thought to
be under control
\cite{Bennett:2003, Tegmark:2003a, Gaztanaga:2003, Efstathiou:2003b}.}.
%[19-22].
The probability of getting the observed CMB
amplitudes depends on the method of evaluation, but it seems
that in about 95\% of its realisations, the
`concordance model' does not reproduce the observed
CMB spectrum \cite{Efstathiou:2003b, Efstathiou:2003a}.
Another phrasing is that the model is ruled out at about
2$\sigma$ level. (In \cite{Gaztanaga:2003} the
discrepancy was evaluated to be much lower, about 70\%, or
1$\sigma$.) The second motivation is not observational
but theoretical.

\paragraph{The coincidence problem.}

The most unattractive feature of the `concordance model' is
the coincidence problem: why has
the acceleration started in the recent past? Or, to phrase
it differently, why has the energy density of dark energy become
comparable to the energy density of matter only recently?
There are three possible answers to this question.

The first possibility is that this is just a coincidence. If the
theory of quantum gravity determines the unique value
of vacuum energy, perhaps this simply happens to be of the
order of the matter energy density today. Since
$(\rho_{\textrm{m}})^{1/4}\approx 10^{-3} \textrm{eV}\approx (\textrm{TeV})^2/M$,
where $M$ is the (reduced) Planck mass, this may not be unreasonable
\cite{Arkani-Hamed:2000}. This is also the scale of neutrino
mass splittings, so there might be a relation \cite{Caldi:1999}.

The second possibility is that there is an anthropic reason. If
various vacua with different vacuum
energies are realised in different parts of the universe, then
the value of vacuum energy in the part of the universe we observe
is naturally so low as not to prevent the formation of galaxies
\cite{Weinberg:1987}. Also, it
will naturally not be so negative as to cause the universe to
collapse very early. This argumentation provides only a window of
values, and one then needs to have some principle or a
specific model to end up with the value apparently 
observed today, or at least to narrow the window sufficiently
\cite{Kallosh:2002}.

The third possibility is that there is a dynamical reason for the
acceleration to have started recently. If so, then this is presumably
related to the dynamics observed in the recent universe
(in principle the dark energy component can of course have
its own dynamics, only weakly related to those of the visible universe).
The important events in recent cosmic history, meaning within the
latest few thousand redshifts, are transition from radiation to matter
domination at around z $=3500$, radiation-matter decoupling at around
z $=1088$ and the growth of structure and related
phenomena at around z $\sim10$ and below.

Models where the dark energy component is sensitive to the
transition from radiation to matter domination have been
constructed \cite{tracker}. In these models the contribution
of the dark energy tracks the radiation density during the
radiation dominated era, and starts to rise after the transition
to the matter dominated era. However, one still generically has
to explain why the dark energy component has started to dominate
at a redshift of at most a few, and not earlier or later. Since
the matter-radiation equality and the start of the acceleration
are far away both in time and in redshift, one could naturally
have the dark energy dominate much earlier or much later.

As for the radiation-matter decoupling, it is nearer to the
start of the acceleration than radiation-matter equality, but does
not seem provide a promising trigger mechanism.

Structure formation occurs around the same redshift as
the acceleration starts, and so the possibility that the
acceleration is related to the growth of inhomogeneities
in the universe seems natural. One way to implement this is
to use the growth of inhomogeneities as a trigger for a dark energy
component \cite{cosmon}. However, one can also look at the effect
of the growth of structure itself, rather than introducing new
fundamental physics that is sensitive to structure formation
(outside the fitting problem to be discussed, this has been
suggested in \cite{Schwarz:2002}).

\paragraph{The fitting problem.}

The cosmological observations leading to the conclusion that
there is a dark energy component have been interpreted in the
context of a homogeneous and isotropic model for the universe.
The reasoning is that since the universe appears to be homogeneous
and isotropic on large scales\footnote{Though it has been argued
that the homogeneity has in fact not been established, on the basis
that the standard statistical tools used to measure deviation
from homogeneity assume homogeneity on large scales
\cite{Pietronero:1999}.}, taking the metric and the
energy-momentum tensor to be isotropic and homogeneous
should be a good approximation.

In the usual approach, one {\it first} takes the average of the metric
and the energy-momentum tensor, and then plugs these averaged
quantities into the Einstein equation. Observables such as the
expansion rate are then calculated from this equation.
Physically, the correct thing to do is to plug the full
inhomogeneous metric and energy-momentum tensor into the
Einstein equation, and {\it then} take the average. Also,
observables should be expressed directly in terms of the
inhomogeneous metric and sources and then averaged.

Since the Einstein equation is non-linear, the equations
for the quantities which have been averaged before plugging
them in (that is, the usual Friedmann-Robertson-Walker
equations) will in general not be the same as the average
of the equations for the inhomogeneous quantities.
We may equivalently say that the averaged quantities do not
satisfy the Einstein equation. This is the fitting problem discussed
in \cite{Ellis}. When one fits the parameters of a
Friedmann-Robertson-Walker model to observational data, is one
fitting the right model? The difference between the equations for
the quantities averaged beforehand and the average of the
equations for the real inhomogeneous quantities is also known
as backreaction.

The fitting problem has been approached from two directions.
One may try to solve the full problem to obtain the equations
satisfied by the averaged quantities, without assuming a given
background
\cite{Noonan:1984, Carfora, Buchert1, Buchert2, Takada:1999, averaging}.
%[31-36].
A more modest approach is to assume a homogeneous
and isotropic background and study the effect of perturbations
on this background
\cite{Futamase, Bildhauer:1991, Zotov, Tsamis, Mukhanov, Russ:1996, Seljak:1996, Boersma:1997, Nambu:2000a, Nambu:2000b, Li:2001, Abramo, Nambu:2002, Geshnizjani:2002, Brandenberger:2002, Finelli:2003, Geshnizjani:2003}.
%[37-53].
(A brief overview of
some of the averaging procedures that have been used is given in
\cite{Stoeger:1999}.) Sometimes the term backreaction is used to refer
only to the second, more limited, approach.

The mathematical problem of obtaining the average metric and the
equations satisfied by the average quantities has not been solved,
though some progress has been made \cite{Buchert1, Buchert2}. The issue of
the metric is particularly complicated, since in order to integrate
a tensor one has to parallel transport its components with respect
to some background, but this background is precisely what one is
trying to determine. From a physical point of view one might expect
that when deviations from homogeneity and isotropy are small,
the homogeneous and isotropic metric should be a good description.
However, the study of backreaction on inflationary backgrounds
has shown that even this is not necessarily true, since
the large number of perturbative modes can compensate for
their small amplitude
\cite{Tsamis, Mukhanov, Nambu:2000b, Li:2001, Abramo, Nambu:2002, Geshnizjani:2002, Brandenberger:2002, Finelli:2003, Geshnizjani:2003}.
%[40, 41, 46-53].

We will take the more modest approach. We assume that there is
a given homogeneous and isotropic background with perturbations
on it and that both satisfy the Einstein equation. We will then
study the effect of these perturbations on the local
expansion rate and see how it differs from the
background expansion rate.

The most straightforward way to consider the impact of
perturbations is to expand the equations in a perturbative series
and solve them in a consistent manner. For cosmological
perturbations this is an involved task. The second order solutions
that are known \cite{Finelli:2003, Acquaviva:2002, Maldacena:2002}
have been built order by order, assuming that the equations for
higher order terms have no impact on the equations for lower order
terms, so they are not fully consistent second order calculations.

We will not solve the second order equations, but will
simply assume the background and the perturbations to be
given by first order perturbation theory. Obviously, if
the impact of the perturbations on the background turns
out to be large, this is no longer a good approximation, and
a consistent second order calculation would be needed.

In section 2 we calculate the local expansion rate for
an observer in a perturbed FRW universe. We evaluate this
for the Einstein-de Sitter case and take the average. We find
a non-vanishing correction to the expansion rate from the
perturbations. In section 3 we discuss the relation to dark
energy and summarise our results.

\section{The backreaction calculation}

\subsection{The local expansion rate}

\paragraph{The metric and the Einstein equation.}

We are interested in the expansion rate measured by a
comoving observer. We will expand this observable in terms
of the perturbations around homogeneity and isotropy, and take the
average. Our approach closely follows that of \cite{Geshnizjani:2002}.

We take the homogeneous and isotropic background spacetime
to be spatially flat. We take the source to be a
single fluid with no anisotropic stress, and we will not consider
vector or tensor perturbations. To first order in perturbations,
the metric can then be written as \cite{Mukhanov:1992}
\bea \label{metric}
  ds^2 = -(1+2\Phi(t,\bx)) dt^2 + (1-2\Phi(t,\bx))\, a(t)^2 d\bx^2 \ .
\eea

The perturbation $\Phi$ coincides with a gauge-invariant quantity
in first order perturbation theory, and is identified as the
gravitational potential in the Newtonian limit. We choose the
background scale factor to be normalised to unity today, $a(t_0)=1$.
An overdot will be used to denote derivative with respect to the
time $t$. Note that because of the perturbations, $t$ is not the proper
time measured by a comoving observer, and one has to be careful to
recast the time-dependence of observables in terms of
the proper time, as emphasised in \cite{Geshnizjani:2002}.

The Einstein equation reads
\bea \label{einstein1}
  G_{\mu\nu} = \frac{1}{M^2} T_{\mu\nu} = \frac{1}{M^2}\left( (\rho + p ) u_{\mu} u_{\nu} + p g_{\mu\nu} \right) \ ,
\eea

\noindent where $M=1/\sqrt{8 \pi G_{\textrm{N}}}$ is the (reduced)
Planck mass, $\rho$ and $p$ are the energy density and pressure of matter,
respectively, and $u^{\mu}$ is the velocity of the matter fluid,
with $u_{\mu} u^{\mu}=-1$.

\paragraph{The expansion rate.}

The observable of interest, the expansion rate measured by an
observer comoving with the matter fluid, is given by
\bea \label{thetadef}
  \theta(t,\bx) = u^{\mu}_{\ \, ;\mu} \ ,
\eea

\noindent where ; stands for the covariant derivative.

In order to evaluate $\theta$ for a typical comoving observer,
we will expand to second order in $\Phi$ and take the average. For
this purpose, let us look at the $0i$-component of \re{einstein1}:
\bea \label{einstein2}
  G^{0i} = \frac{1}{M^2} (\rho + p) u^{0} u^{i} = \frac{1}{3} (4 G_{\mu\nu} u^{\mu} u^{\nu} + G_{\mu\nu} g^{\mu\nu} ) u^{0} u^{i} \ ,
\eea

\noindent where we have used the Einstein equation \re{einstein1}
again. Writing $G_{\mu\nu}$ in terms of the metric, we have an
iterative equation from which $u^{\mu}$ can be solved
to any desired order in $\Phi$. Given the initial condition that for the
background spacetime $u^{\mu}=(1,\bfm{0})$, we get to second order
\bea \label{u}
  u^{0} &\simeq& 1 - \Phi + \frac{3}{2}\Phi^2 + \frac{1}{2}\frac{1}{a^2 \Hdot^2}\pat_i (\Phidot + H\Phi)\pat_i (\Phidot + H\Phi) \el
  u^{i} &\simeq& \frac{1}{a^2 \Hdot}\pat_i (\Phidot + H\Phi) + \frac{1}{a^2 \Hdot}( \Phidot\pat_i\Phi + 5 \Phi\pat_i\Phidot + H\Phi\pat_i\Phi ) \el
  && + \frac{1}{a^2 \Hdot^2}\pat_i (\Phidot + H\Phi) \left( \Phiddot + H\Phidot + \frac{1}{a^2}\nabla^2\Phi \right) \ ,
\eea

\noindent where $H=\adot/a$ is the background expansion rate.

It is noteworthy that $u^{\mu}$ contains terms with more than
two derivatives (specifically spatial derivatives).
This perhaps surprising feature can be understood from
\re{einstein2} as follows. Let us assume
that $u^{i}$ contains exactly one spatial derivative
(a non-zero $u^{i}$ must have at least one spatial derivative
to give the index). Then the right-hand side of \re{einstein2}
contains products of $G_{ij}$ and three powers of $u^{i}$, 
which in general contain five spatial derivatives (and do not
cancel). But this
makes the equation inconsistent, since the left-hand side only
contains terms with at most two derivatives. Obviously,
the conclusion also holds for two, or any other finite number of,
spatial derivatives in $u^{i}$: the only possibilities for the
number of derivatives in $u^{\mu}$ that are consistent with
\re{einstein2} are zero and infinite. So, the number of derivatives
in an iterative solution for $u^{\mu}$ depends on the power of
$\Phi$ at which the iteration is stopped. Since the derivative terms
arise by combining components of $G_{\mu\nu}$ algebraically,
there are no terms with more than two derivatives, either
spatial or temporal, acting on a given $\Phi$. Therefore
the maximum number of derivatives at order $\Phi^N$ is $2 N$.

Plugging the expression \re{u} for $u^{\mu}$ into \re{thetadef}
we have, to second order,
\bea \label{theta}
  \fl \theta \simeq 3 H - 3 (\Phidot + H\Phi) - 3\Phi\Phidot + \frac{9}{2} H \Phi^2 + \frac{3}{2}\frac{H}{a^2 \Hdot^2}\pat_i (\Phidot + H\Phi)\pat_i (\Phidot + H\Phi) \el
  \!\!\! - 2 \frac{1}{a^2 \Hdot}\pat_i(\Phidot + H\Phi)\pat_i\Phi +\pat_t\left( \frac{1}{2}\frac{1}{a^2 \Hdot^2}\pat_i (\Phidot + H\Phi)\pat_i (\Phidot + H\Phi) \right) + \pat_i u^i  \ ,
\eea

\noindent where the total gradient $\pat_i u^i$ has not been
written explicitly. Since $u^{\mu}$ contains
(at second order) terms with four derivatives, $\theta$ contains
terms with five derivatives. The next to last term has 
two spatial and three temporal derivatives and the
last term, $\pat_i u^i$, has a contribution with two spatial
and three temporal derivatives and a contribution with
four spatial derivatives and one temporal derivative.

\paragraph{The proper time.}

In order to find the physical expansion rate, we should recast
\re{theta} in terms of the proper time $\tau$ of a comoving
observer. The derivative in the direction orthogonal to the
hypersurface defined by the velocity $u^{\mu}$ is
$\pat_{\tau}=u^{\mu}\pat_{\mu}$. From the condition
$\pat_{\tau}\tau=1$ we obtain, using \re{u}, an iterative
equation for $\tau$. Given the initial condition that for
the background spacetime $\tau=t$, we get to second order
\bea \label{tau}
  \fl \tau \simeq t + \int^{t} \rmd t' \bigg( \Phi - \frac{1}{2}\Phi^2 -  \frac{1}{2}\frac{1}{a^2 \Hdot^2}\pat_i (\Phidot + H\Phi)\pat_i (\Phidot + H\Phi) \el
  - \frac{1}{a^2 \Hdot}\pat_i (\Phidot + H\Phi) \int^{t'} \rmd t''\pat_i\Phi \bigg) \ .
\eea

If we neglected the gradient terms, we would get (to all orders in $\Phi$)
\mbox{$\tau=\int\rmd t \sqrt{|g_{00}|}$}, in agreement with
\cite{Geshnizjani:2002}, where gradients were dropped.

From \re{u}, \re{theta} and \re{tau} we can calculate the expansion rate
in terms of the proper time. For general functions $a$ and
$\Phi$ the expression is cumbersome and not very illuminating.
However, for the Einstein-de Sitter universe things simplify
considerably.

\paragraph{The Einstein-de Sitter universe.}

As discussed earlier, we take $a$ and $\Phi$ from the first
order formalism and calculate $\theta$ to second
order with these expressions. If the effect of the
perturbations on $\theta$ is large, then this approach
has reached its limit of validity, and a consistent second order
calculation would be needed.

The matter is taken to be pure cold dark matter, $\Omega_{\textrm{cdm}}=1$.
When relevant, we will mention what effect a realistic
baryon content of $\Omega_{\textrm{b}}=0.05$ \cite{Spergel:2003, Cyburt:2003}
would have; at our level of approximation, the difference is minimal.
The perturbations are taken to be purely adiabatic,
Gaussian (with zero mean), and to have a scale-invariant spectrum,
$n=1$. We will also consider only the growing mode of the perturbations.

The background expansion rate is given by the FRW solution for
pressureless matter, so $a=(t/t_0)^{2/3}$ and $H=2/(3 t)$.
We take the value of $H$ today to be $H_0=h$ 100 km/s/Mpc,
with $h=0.7$; our results are not sensitive to the precise
value of $h$. For $a=(t/t_0)^{2/3}$, the local expansion rate is,
from \re{u}, \re{theta} and \re{tau},
\bea \label{thetamatter}
  \fl \theta \simeq 3 H_{\tau} - 3 (\Phidot + H\Phi) + 3 H\frac{1}{t}\int\rmd t\Phi - 3\Phi\Phidot + \frac{9}{2} H \Phi^2 - 3 H \left(\frac{1}{t}\int\rmd t\Phi\right)^2  \el
  - \frac{3}{2} H \frac{1}{t}\int\rmd t\Phi^2 + \frac{2}{3}\frac{1}{a^2 H^3} \pat_i (\Phidot + H\Phi)\pat_i (\Phidot + H\Phi) \el
  + \frac{4}{3}\frac{1}{(a H)^2} \pat_i (\Phidot + H\Phi)\pat_i\Phi + \frac{2}{9}\pat_t\left( \frac{1}{(a H)^2 H^2}\pat_i (\Phidot + H\Phi)\pat_i (\Phidot + H\Phi) \right) \el
  - \frac{2}{3} H\frac{1}{t}\int\rmd t \frac{1}{(a H)^2 H^2}\pat_i (\Phidot + H\Phi)\pat_i (\Phidot + H\Phi) \el
  + 2 H\frac{1}{t} \int^{t} \rmd t' \left( \frac{1}{(a H)^2}\pat_i (\Phidot + H\Phi) \int^{t'} \rmd t''\pat_i\Phi \right) + \pat_i u^{i} \ ,
\eea

\noindent where we have defined $H_{\tau}=2/(3\tau)$, and
$\pat_i u^{i}$ has again not been written explicitly.
The terms after $3 H_{\tau}$ are the backreaction contribution.

\subsection{The average expansion rate}

\paragraph{Taking the average.}

To evaluate the backreaction, we should take the
average of \re{thetamatter} over the hypersurface of constant
$\tau$. The backreaction has been expressed in terms of the
background coordinates $t$ and $x^i$. We have to rewrite it
in terms of the proper time $\tau$
and spatial coordinates orthogonal to $\tau$, denoted 
by $y^i$. We also have to take into account the
integration measure on the hypersurface of constant
$\tau$. After a somewhat lengthy calculation, we get
\bea \label{average}
  \fl <\theta> \simeq 3 H_{\tau} - \pat_{\tau}<3 \Phi +2 \phi + \frac{2}{3}\frac{1}{(a_{\tau} H_{\tau})^2}\nabla^2\Phi >_0 \el
  - \frac{1}{2} \pat_t\bigg( < 3\Phi + 2\phi + \frac{2}{3} \frac{1}{(a H)^2}\nabla^2\Phi >_0 \bigg)^2 \el
  + 2 \frac{1}{H}\pat_t^2 <\phi\Phi>_0 + 14 \pat_t<\phi\Phi>_0 + \pat_t<\phi^2>_0 - \frac{3}{2}\pat_t<\Phi^2>_0 \el
  + \frac{3}{2} H\frac{1}{t}\int\rmd t t\pat_t<\Phi^2>_0 - \frac{4}{9}\frac{1}{H} \pat_t^2 \left( \frac{1}{(a H)^2}<\pat_i\phi \pat_i\Phi>_0 \right) \el 
  - \frac{24}{9} \pat_t \left( \frac{1}{(a H)^2}<\pat_i\phi \pat_i\Phi>_0 \right) + \frac{2}{3}\pat_t\left( \frac{1}{(a H)^2}<\pat_i\Phi \pat_i\Phi>_0 \right) \el
  + \frac{2}{9} \pat_t \left( \frac{1}{(a H)^2 H^2}<\pat_i(\Phidot+H\Phi) \pat_i(\Phidot+H\Phi)>_0 \right) \el
  + 2 H\frac{1}{t}\int\rmd t t\pat_t \left( \frac{1}{(a H)^2}<\pat_i\Phi \pat_i\Phi>_0 \right) \el
  + \frac{2}{3} H\frac{1}{t} \int\rmd t t \pat_t \left( \frac{1}{(a H)^2 H^2}<\pat_i(\Phidot+H\Phi) \pat_i(\Phidot+H\Phi)>_0 \right) \el
  + \frac{2}{9} \frac{1}{(a H)^2}< \pat_i \bigg[ 8\Phidot\pat_i\Phi - 6\Phi\pat_i\Phidot + 15 H\Phi\pat_i\Phi  + 2 \frac{1}{H}\Phidot\pat_i\Phidot \el
  + 2\frac{1}{H^2}\Phiddot\pat_i(\Phidot + H\Phi) + 2 \frac{1}{H} \phi \pat_i \Phiddot + 10 \phi\pat_i\Phidot -4 H\phi\pat_i\Phi\bigg]>_0 \el
  + \frac{4}{9}\pat_t\left(\frac{1}{(a H)^4} <\pat_i\left[ \nabla^2 \Phi\pat_i\Phi \right]>_0\right) \ ,
\eea

\noindent where we have defined $\phi=t^{-1}\int\rmd t \Phi$
and $a_{\tau}=(\tau/t_0)^{2/3}$. (Note that when $\Phidot=0$,
we have $\phi=\Phi$.) Here the arguments
of $\Phi$ and its derivatives are understood to be
$(\tau,\by)$ rather than $(t,\bx)$, and $<A>_0$ denotes
$(\int\rmd ^3 y)^{-1}\int\rmd ^3 y A(\tau, \by)$. Since the
calculation is only to second order, we have substituted
$t$ for $\tau$ in the terms with two powers of $\Phi$.

The backreaction terms can be divided into three groups:
linear terms, quadratic terms which are not total gradients
and quadratic terms which are total gradients (and
therefore reduce to boundary terms). It is noteworthy that
all terms apart from the first total gradient term
are total time derivatives of dimensionless expectation values:
for example, there are no terms of the form $H (a H)^{-2}<\pat_i\Phi\pat_i\Phi>_0$.

\paragraph{The linear terms.}

It might seem that the average of the linear backreaction
terms must be zero since the perturbations
are assumed to be Gaussian (with zero mean). However, we have not
specified the hypersurface with respect to which this holds.
Are the perturbations Gaussian with respect to the (unphysical)
hypersurface of constant $t$ of the background spacetime, or
with respect to the (physically meaningful) perturbed
hypersurface of constant $\tau$? Obviously, if the perturbations
are distributed according to Gaussian statistics on the
background spacetime, they are not Gaussian with respect to
the perturbed spacetime, and vice versa.

The situation is ambiguous because this question only arises at
second order, and we are plugging in perturbations from first
order perturbation theory. The issue would probably have to be
resolved by a consistent second order calculation.
Note that this non-Gaussianity related to the choice of
hypersurface is distinct from the intrinsic non-Gaussianity of
second order perturbations found in
\cite{Acquaviva:2002, Maldacena:2002}.

It seems more physically meaningful to take the perturbations
to be Gaussian with respect to the hypersurface of constant $\tau$.
However, we keep in the simple approximation of using first
order results for the perturbations, and take them to be
Gaussian with respect to the
background. Then the linear terms, and the squares of linear terms,
in \re{average} vanish. Under another assumption about the
statistics, this may not be true.

\paragraph{The quadratic non-total gradient terms.}

In the Einstein-de Sitter universe, the non-decaying mode of
$\Phi$ is constant in time, $\Phidot=0$, in first order
perturbation theory. The expansion rate \re{thetamatter}
then simplifies to
\bea \label{theta1}
  \theta \simeq 3 H_{\tau} + \frac{118}{45}\frac{1}{a^2 H}\pat_i\Phi\pat_i\Phi + \pat_i u^i \ .
\eea

It is noteworthy that all non-gradient correction terms have
disappeared: only gradients contribute to the backreaction,
in agreement with \cite{Geshnizjani:2002}. The average expansion
rate \re{average} simplifies to (neglecting the linear terms and their
squares),
\bea \label{average1}
  \fl <\theta> \simeq 3 H_{\tau} \left( 1 - \frac{22}{135}\frac{1}{(a H)^2}<\pat_i\Phi\pat_i\Phi>_0  + \frac{22}{27}\frac{1}{(a H)^2} <\pat_i (\Phi\pat_i\Phi)>_0 \right. \el
   \left. + \frac{8}{27}\frac{1}{(a H)^4} <\pat_i \left(\nabla^2\Phi\pat_i\Phi\right)>_0 \right) \ .
\eea

Postponing discussion of the total gradient terms,
let us evaluate the first correction term:
\bea \label{back1}
  \frac{1}{(a H)^2} <\pat_i\Phi\pat_i\Phi>_0 &=& \frac{1}{(a H)^2}\int_0^{\infty}\frac{\rmd k}{k} k^2 \Delta^2_{\Phi}(k) \el
  &\approx& \frac{9}{4} \frac{(a H)^4}{(a H)^2} \int_0^{\infty}\frac{\rmd k}{k} k^{-2} \Delta^2_{\delta}(k,a) \el
  &=& \frac{9}{4} \frac{1}{(a H)^2} \int_0^{\infty}\frac{\rmd k}{k} A^2 k^{2} T(k)^2 \el
  &\approx& 8\cdot 10^{-5} a \ ,
\eea

\noindent where $\Delta_{\Phi}^2$ and $\Delta_{\delta}^2$
are the power spectra of the metric and density perturbations,
respectively, and we have used
$\Phi_{\bk}\approx -3 (a H)^2 \delta_{\bk}/(2 k^2)$
(the low-$k$ part of the spectrum where this is not a good approximation
gives negligible contribution).
The density power spectrum is taken to be
$\Delta_{\delta}^2(k,a)=A^2 k^4 T(k)^2/(a H)^4$,
where $A=1.9\cdot 10^{-5}$ and $T(k)$ is the CDM transfer function,
for which we use the BBKS fitting formula \cite{Bardeen:1986}.
Taking into account a baryon contribution of $\Omega_{\textrm{b}}=0.05$
with the shape parameter $\Gamma=h\exp( -\Omega_{\textrm{b}} )$
\cite{Weinberg:2002} would only change the prefactor from 8 
to 7 in \re{back1}.

The magnitude of the effect can be understood as follows.
Due to the transfer function, the main contribution comes from
around $k_{\textrm{eq}}$, so a rough estimate is
$H_0^{-2} <k^2\Phi^2>\sim (k_{\textrm{eq}}/H_0)^2 <\Phi^2> \sim (150)^2 (2\cdot 10^{-5})^2 \sim 10^{-5}$. Note that the backreaction is enhanced by the 
large factor $(k_{\textrm{eq}}/H_0)^2$ and so could be large even with
$\Phi$ much below unity.

Taking into account the numerical factor from \re{average1},
we get $-1\cdot 10^{-5}$ for the relative correction.
The negative sign may seem surprising,
since from \re{theta1} it might appear that the backreaction
definitely increases the expansion rate. This is true when
evaluated over the background surface of constant $t$, but taking
into account the perturbations in the hypersurface changes the
sign. To appreciate the importance of the choice of hypersurface, it
may be helpful to note that evaluating the expansion rate $3 H_{\tau}$
over the background hypersurface of constant $t$ gives an apparent
backreaction contribution. The choice of hypersurface has been
discussed in \cite{Geshnizjani:2002, Geshnizjani:2003}.

The backreaction seems to increase proportional to the 
scale factor $a$, and would seem to be
more important in the future, eventually dominating the
expansion rate. However, this is not true, because
the linear regime of perturbations does not extend to
infinitely small scales, due to the process of structure formation.
We will take this into account by introducing a time-dependent
cut-off, denoted by $k_{\textrm{L}}$, at the scale at which the mean square
of the density perturbations becomes unity, $\sigma^2=1$\footnote{We should also introduce a lower cut-off
at the horizon scale $(a H)/2$, to take into account that
the average is properly over the horizon volume and not over
all space. However, for the terms considered here, the effect
of this cut-off is negligible.}. The end
of the linear regime of density perturbations is today at
around $k_{\textrm{L}}=0.1$ Mpc$^{-1}$. Putting a cut-off in the integral
\re{back1} at this $k_{\textrm{L}}$ only changes the prefactor from 8 to
2 (at this level of accuracy, the baryons make no difference).
However, since we are cutting the perturbations off at some
time-dependent scale in momentum space, the
perturbations in position space are no longer time-independent.
Therefore, the simple result \re{average1} applies only
before the formation of bound structures, that is, before
$\sigma^2=1$ on any scale.

Note that we are applying the cut-off to the perturbations
themselves, not to the integration range. This seems to be
more physically correct for the high-$k$ cut-off (though not
for the horizon cut-off, were we to introduce one). If we applied
the cut-off to the integration range instead, we would
have $\Phidot=0$ and the simple result \re{average1} would still
hold\footnote{Note that in \re{average} the integration range
has been assumed not to depend on time.}. The treatment of the onset
of non-linearity will be discussed in more detail in section 3.

For a realistic transfer function, it is not possible to
calculate the full backreaction \re{average} analytically in
the presence of a time-dependent cut-off in momentum space.
However, we can determine the time behaviour of the
backreaction by looking at its asymptotic behaviour.

The asymptotic future value of \re{average} can be found by
a simple dimensional argument. There are three scales in the
problem, $a H$, $k_{\textrm{L}}$ and (from the transfer function)
$k_{\textrm{eq}}$;  the scale factor $a$ enters only via these
scales. Therefore the non-gradient
expectation values such as $<\Phi^2>_0$ must be some function of
$k_{\textrm{L}}/(a H)$ and $k_{\textrm{eq}}/(a H)$, and the
gradient expectation
values such as $<\pat_i\Phi\pat_i\Phi>_0$ must be $(a H)^2$
(or $(a H)^2 H^2$ in the case of expectation values involving
$\Phidot + H\Phi$) times some function of the same two variables. 

In the future, structure formation will have proceeded so that
all scales smaller than $1/k_{\textrm{eq}}\approx$ 30 Mpc will have gone
non-linear. The CDM transfer function $T(k)$ is essentially unity
for scales much smaller than $k_{\textrm{eq}}$, and for $T(k)=1$
the condition $\sigma^2=1$ gives
$k_{\textrm{L}}=\sqrt{2/A} (a H)\propto a H$\footnote{For simplicity, we are
using a top hat in momentum space as the window function.}.
The scale $k_{\textrm{eq}}$ has disappeared,
and $k_{\textrm{L}}$ is proportional to $a H$.
It follows that the non-gradient
expectation values are constant, and the gradient ones are
simply constants times $(a H)^2$ (or $(a H)^2 H^2$, as appropriate).
The terms we are considering
contribute to the average expansion rate \re{average} only
via total time derivatives of dimensionless expectation values.
Therefore, the backreaction vanishes.

So, the backreaction from the quadratic terms which
are not total gradients behaves
as follows. Before $\sigma^2=1$ on any scale, the backreaction is
given by \re{back1}. It slows down the expansion rate, its relative
contribution rises with time and is of the order $10^{-5} a$.
When the limit $\sigma^2=1$ is reached, two things happen.
First, the integral \re{back1} receives a
cut-off at $k_{\textrm{L}}$, which decreases as $a$ increases and larger
scales become non-linear. Second, time derivatives of the
perturbations start contributing. The net effect
is to reduce the magnitude of the backreaction, so
that it is asymptotically zero in the future.

It should be noted that for the scale-invariant spectrum
that we have considered, the limit $\sigma^2=1$ is actually
reached at all times on sufficiently small scales. Even taking
into account collisional damping and free streaming, the limit
is reached quite early for a pure adiabatic spectrum of CDM
perturbations. In \cite{Green:2003}, structure formation was
estimated to start at z $\approx40-60$ in the case of supersymmetric CDM.
For light dark matter, the power spectrum would be more damped,
and the limit would be reached later \cite{Boehm:2003}.

The conclusion of asymptotically vanishing backreaction depends
on the absence of a scale other than $a H$: if there is another
scale present, the backreaction will in general not vanish. In
particular, the backreaction will be non-zero if the power spectrum
of density perturbations is not scale-invariant. If the spectrum
is given by a power-law with a constant spectral index $n$, the
integrand in \re{back1} will be multiplied by $(k/k_P)^{n-1}$,
where $k_P$ is some constant scale. As discussed above, in the
future the transfer function will be unity for all modes in the
linear regime. Then the integral \re{back1} will be proportional
to $k_{\textrm{L}}^{n-1}\propto (a H)^{n-1}$. Therefore, the
total time derivatives in \re{average} will not vanish, and the
relative backreaction will be proportional to
$(a H)^{n-1}\propto a^{(1-n)/2}$. For a red
spectrum, $n<1$, the backreaction will grow and eventually
dominate the expansion of the universe. The impact on
the expansion rate will be discussed in more detail in section 3.

\paragraph{The quadratic total gradient terms.}

The backreaction terms previously considered could be readily
evaluated using the standard methods of cosmological
perturbation theory. The remaining total gradient terms
are more problematic. In the standard treatment,
cosmological perturbations are assumed to be periodic on some large
scale, so that one can decompose them as a Fourier series.
It is assumed that the periodicity has no impact on observables
as long as the periodicity scale is large enough. However,
the average of a total gradient yields a boundary term
which is of course sensitive to boundary conditions: the
periodicity forces the average of a total gradient to
vanish. (The conditions imposed by the existence of a
Fourier transform are less straightforward, but they also
imply the vanishing of averages of total gradients.)
There is no obvious physical reason for the quadratic total
gradient terms in \re{average} to vanish, so this seems
to be just a mathematical artifact.

As noted earlier, we should properly take the average over only
the present horizon, while the periodicity scale should be
taken to be much larger than the horizon. Therefore, the proper
way to evaluate the total gradient terms in the standard
treatment of cosmological perturbations is to take a box much
bigger than the horizon and consider the effect of these terms
on horizon-sized subsamples of space. If the hypothesis that
the periodicity has no impact on observables (for a sufficiently large
box) is correct, then the result should be the same that we would
get by evaluating the total gradient terms with realistic
boundary conditions, not imposing periodicity.
Note that the time development of the backreaction terms (and
therefore the apparent equation of state, to be discussed in
section 3), is fixed by the time-development of the perturbations,
so that only the magnitude remains to be determined.

Such a calculation has been done in a different, Newtonian,
backreaction formalism, where backreaction vanishes
completely for periodic boundary conditions \cite{Buchert2}.
The box was taken to be of about the horizon size, and it was
found that the backreaction can have a substantial impact
on the expansion rate on scales smaller than about 50 Mpc,
either increasing or decreasing it.

We will not here embark on such a numerical computation, but will
naively estimate the magnitude of the total gradient terms.
The total gradient terms in \re{average}
have the form $<\pat_i(f\pat_i g)>_0=<\pat_i f\pat_i g>_0 + <f\nabla^2 g>_0$.
Evaluated over all space, the two parts cancel (in the standard
treatment of perturbations). Evaluated over a volume
smaller than the box size, the cancellation will not be perfect
(or the terms may even have the same sign). We will therefore
crudely estimate the magnitude of the backreaction by looking at the
parts evaluated individually over all space.

The first total gradient term in \re{average}
contains two spatial derivatives and is thus similar to the
terms previously evaluated. When this term does not average to
zero, it is expected to modify the numerical
coefficients of the previously considered gradient terms
by factors of at most order one, possibly changing their sign.
It could also make the asymptotic value of the backreaction
different from zero.

The second total gradient term in \re{average} is
qualitatively different from any others in that it
contains four spatial derivatives. Since
\bea \label{back2}
  \frac{1}{(a H)^4} <\nabla^2\Phi\nabla^2\Phi>_0 = \frac{9}{4}<\delta^2>_0 \equiv \frac{9}{4}\sigma^2 \ ,
\eea

\noindent the first part of the backreaction term is simply
$-\pat_t \sigma^2$ and the second is obviously $\pat_t \sigma^2$.
Before the limit $\sigma^2=1$ is reached on any scale,
the first part of the backreaction term (relative to $3 H_{\tau}$)
is given by $-\frac{2}{3}\,\sigma^2$ and the second by
$\frac{2}{3}\,\sigma^2$,
since $\sigma^2\propto a^2$. After the limit $\sigma^2=1$
is reached, we should introduce a cut-off at $k_{\textrm{L}}$, meaning that
$\sigma^2$ saturates at unity and both parts of the backreaction
vanish independently, $\pat_t \sigma^2=0$.

So, the contribution of the total gradient term with four
spatial derivatives grows like $a^2$ until the limit
$\sigma^2=1$ is reached, at which point it drops ro zero.
(The expansion rate thus jumps discontinuously when
$\sigma^2$ reaches unity, due to the unrealistic sharp cut-off
at $k_{\textrm{L}}$.) The average over all space is always zero, and so
the backreaction before the start of the formation of bound
objects will lead to to slower expansion in some regions
and to faster expansion in others. A reasonable estimate of the
relative contribution to the expansion rate might be of the
order $\frac{2}{3}\,\sigma^2$ or so, though in some regions it will be much
smaller or larger. The probability distribution of magnitudes
should be properly evaluated, for example by numerical computation.
Also, the conclusions here are sensitive to the sharp cut-off at
$k_{\textrm{L}}$. For example, clearly the physical expansion rate
will not jump discontinuously when the limit $\sigma^2=1$ is
reached, so the backreaction could be large
for at least some time afterwards.

\section{Discussion}

\paragraph{The effective dark energy.}

To appreciate the impact of backreaction, let us look at
the Hubble law. Let us parametrise the quadratic backreaction 
terms in \re{average} as follows:
\bea \label{hubble}
 \fl \left( \frac{1}{3}<\theta> \right)^2 &\simeq& H_{\tau}^2 ( 1 + \lambda_1 a^{m_1} + \lambda_2 a^{m_2})^2 \el
  &=& H_{\tau}^2 + H_{\tau}^2 ( 2 \lambda_1 a^{m_1} + 2 \lambda_2 a^{m_2} + 2 \lambda_1 \lambda_2 a^{m_1+m_2} + \lambda_1^2 a^{2 m_1} + \lambda_2^2 a^{2 m_2} ) \ ,
\eea

\noindent where the subscripts 1 and 2 refer to terms with one
and two powers of $k^2/(a H)^2$, respectively.
The coefficient $\lambda_1$ is of the order $10^{-5}$,
while $\lambda_2$ may give a contribution of order one
before the limit $\sigma^2=1$ is reached, and is expected to
drop to zero afterwards. Before the limit $\sigma^2=1$, we
have $m_1=1$ and $m_2=2$. After the threshold $\sigma^2=1$ is
reached, $m_1$ slowly approaches zero, while $m_2$
is expected to go to zero rapidly. (The power-law behaviour
in \re{hubble} is obviously only valid piecewise.)

To someone fitting the observed expansion rate to the homogeneous
and isotropic FRW equation $(\theta/3)^2=\rho/(3 M^2)$ it
would seem that there is a mysterious energy component which
affects the expansion rate but which is nowhere to be seen.
The apparent equation of state $w$ of the `dark energy'
is easy to determine from the FRW relations
$H^2\propto a^{-3}$, $\rho_{\textrm{de}}\propto a^{-3 (1+w)}$.
These give the equations of state $w=-m/3$, where $m$
is $m_1$, $m_2$, or one of the combinations in \re{hubble}.
Before structure formation, we would have a mixture of
the equations of state $-\frac{1}{3}, -\frac{2}{3}, -1$ and $-\frac{4}{3}$. 
(Note that it is perfectly natural to get an equation of state
which is more negative than $-1$.) Today, the equation
of state would be between $-\frac{2}{3}$ and 0. The
relative dark energy density today would
seem to be $\Omega_{\textrm{de}}=1-1/(1+\lambda_1+\lambda_2)^2$.

What does this imply for the expansion history of the
universe? Neglecting $\lambda_1$ as probably small,
we are left with $\lambda_2$. Early on, before $\sigma^2=1$
on any scale, its relative contribution is related to
$\sigma^2$ and rises like $a^2$ and $a^4$, corresponding to
the mixture of equations of state $-\frac{2}{3}$ and $-\frac{4}{3}$. After
the limit $\sigma^2=1$ is reached, the contribution drops to zero.
This term would induce a period of acceleration if its sign is
positive and its contribution is comparable to that of
dark matter, which could be the case around the beginning
of the formation of bound structures. While the period of acceleration
would drive the Hubble parameter up from the FRW Einstein-de Sitter
value, this sort of an expansion history probably could not account
for the supernova data.

After $\lambda_2$ goes to zero, the only backreaction comes from
$\lambda_1$. The contribution from this term has probably always
been negligible, but for a red spectrum it grows in time. As discussed
earlier, for a constant spectral index $n$ the relative backreaction
will grow like $a^{(n-1)/2}$ in the future, giving a combination
of the equations of state $(n-1)/6$ and
$(n-1)/3$. Even for the quite red spectrum with $n=0.8$ in
\cite{Blanchard:2003}, we would only get a mixture of $w\approx-0.03$
and $w\approx-0.07$. The backreaction would grow very slowly
and eventually cause the Einstein-de Sitter universe
to collapse. (Of course, the approximation that the perturbations
behave according to first order perturbation theory would break down
before that, and a consistent second order calculation would be needed.)

One should be careful with the Hubble law, because the
square of $<\theta>$ is not the same as the average of
$\theta^2$. From \re{u}, \re{tau} and \re{thetamatter}
we find (to second order) the following relation between
the average of the square and the square of the average:
\bea \label{variance}
  <\theta^2> - <\theta>^2 = \pat_{\tau}<\theta> - <\pat_{\tau}\theta> \ .
\eea

The above relation has previously been found (in an exact form)
in the different backreaction formalism of
\cite{Buchert1, Buchert2}. The physical content of
\re{variance} is that the relative change of the integration
measure on the hypersurface of constant proper time
with respect to the proper time gives the
expansion rate. Denoting the integration measure by $J$,
we have $J^{-1}\pat_{\tau}J=\theta$. For example, for a
spatially flat FRW spacetime we have $a^{-3}\pat_t (a^3)= 3 H$.

One should be careful to identify the observable actually
measured by an experiment and compare the theoretical average of that
observable with the experimental data. For the supernova data,
it seems more correct to take the average first and then
square it. Averaging the square instead would lead
to a positive (non-total gradient) contribution of
$\frac{1}{9}\,\sigma^2$ to the square of the Hubble rate before structure
formation (with the equations of state $-\frac{2}{3}$ and $-\frac{4}{3}$),
and a correction of order one afterwards (with the equation
of state $0$). Note that \re{variance} indicates that the
deduced value of $\pat_{\tau}\theta$ is larger if one takes
the average first, in contrast to what happens with $\theta^2$.

As an aside, we note that with the relation $J^{-1}\pat_{\tau}J=\theta$
the average expansion rate can be written in the simple form
\bea \label{J}
  <\theta> = <J>_0^{-1}<J\theta>_0 = \pat_{\tau}(\ln <J>_0) \ .
\eea

The form \re{J} is more suited to a systematic study of higher
order terms than the straightforward calculation used to arrive at
\re{average}. From \re{J} it seems transparent
that the backreaction is given by total time derivatives
of dimensionless expectation values, something that was
not obvious earlier. This may seem to be at odds
with the fact that one of the backreaction terms in
\re{average} did not seem to be a total time derivative.
The resolution is that any term can be written as a total time
derivative simply as $<f>_0$=$\pat_{\tau}\!\!<\int\rmd \tau f>_0$.
Backreaction from most terms in \re{average}
was found to vanish asymptotically because they contained no
such integrals, but this is not true for all terms.
So, while the form of \re{J} looks suggestive,
it in fact contains no information about whether the backreaction
can be written in such a total derivative form as would
vanish asymptotically in the future.

\paragraph{Transition to non-linearity.}

The calculation has been made for perturbations in the
linear regime, and we have treated the onset of non-linearity by
simply introducing a sharp cut-off for the
perturbations at a transition scale $k_{\textrm{L}}$. A more realistic
treatment of the transition would include a description
of the perturbation modes smoothly becoming time-dependent
and breaking away from the general expansion. The effect of
this on the backreaction is not clear, but since the magnitude
of $\Phi$ increases, one would expect the backreaction to increase,
and it could also change sign. Another reason to expect a large
effect is that high-$k$ modes give the main contribution to the
backreaction integrals \re{back1} and \re{back2}.

Backreaction from the non-linear regime has been considered
in \cite{Seljak:1996}, in an approach where the Einstein equation
was averaged, yielding a backreaction term like \re{back1}. The
result for the backreaction from the linear regime of
density perturbations was of the same order, $10^{-5}$,
but with a positive sign. The calculation
was extended into the non-linear regime of density perturbations
using the relation $\Phi_{\bk}\approx -3 (a H)^2 \delta_{\bk}/(2 k^2)$, 
with the result that backreaction from the non-linear regime is
also negligible. However, this result is not reliable.

As discussed above and in section 2, one has to be careful to
identify the correct observable, to recast its time-dependence
in terms of the physical proper time $\tau$ and to take the
average over the hypersurface of constant $\tau$. In \cite{Seljak:1996}
the quantity considered was $(\adot/a)^2$, which does not give
the expansion rate measure by a comoving observer, as seen
from \re{theta}. Also, the time coordinate used was $t$ and the
average was taken over the hypersurface of constant $t$.
The sign difference between the present calculation
and the result of \cite{Seljak:1996} for the linear regime
arises from these factors. (As mentioned after \re{back1}, if
one took the average of \re{theta1} over the hypersurface
of constant $t$, the contribution of the non-total
gradient terms would be positive.)

For the non-linear perturbations, there are two other important
issues. First, the linear relation between
$\Phi_{\bk}$ and $\delta_{\bk}$ is not valid in
the non-linear regime, even though $\Phi_{\bk}$ calculated from this
equation is perturbatively small (since $k\gg a H$).
For example, the gravitational potential inside stabilised
collapsed objects is constant, whereas the relation
$\Phi_{\bk}\propto\delta_{\bk}/a$ would give
$\Phi_{\bk}\propto a^{1/2}$ in the stable clustering approximation.

Second and more relevant, the calculation of \cite{Seljak:1996}
does not take into account
that very non-linear structures have broken away from the expansion. 
We are interested in the locally measured expansion rate, and
should not include contributions from inside the structures whose
relative motion we are considering.
This is not to say that any effect of the gravitational
fields inside bound structures on the average expansion is ruled
out, simply that such effects cannot be captured by the
perturbative formalism of \cite{Seljak:1996} or the present paper.

However, the effect of perturbations in the process of breaking
away from the expansion can be calculated in a perturbative
framework. Such perturbations could significantly change the
results for the apparent dark energy, since their absolute value
rises by orders of magnitude in the process of breaking away.
Apart from boosting the magnitude, one would expect the
growth of the perturbations to also make the equation of state
more negative.

Since the sign of the global backreaction
we calculated for the linear perturbations turned out to
be negative, one might think that increased inhomogeneity
in general globally slows down the expansion.
However, this is not the case. This can be easily seen by
considering the average expansion rate \re{average} with the
power-law behaviour $\Phi\propto a^l$, with constant $l$.
The contribution of the non-gradient quadratic backreaction
terms in \re{average} to the expansion rate is positive
for all $l>0$, while the contribution of the quadratic
(non-total) gradient terms turns out to be positive for $l>\frac{1}{2}$.
This sort of power-law behaviour is not physically relevant
(except possibly piecewise in momentum space), but it shows
that increased inhomogeneity can lead to faster expansion.

The answer that the backreaction of perturbations breaking away
from the general expansion
might give to the dark energy question would be simple: there appears to
be a dark energy component because the observations are fitted
to a model that does not take into account the impact of
inhomogeneities on the expansion rate. This would also naturally
solve the coincidence problem: the backreaction would be small
until perturbations start becoming large.

\paragraph{Conclusion.}

The essence of the results is that there is a non-vanishing
backreaction from inhomogeneities in an Einstein-de Sitter
background, and that its magnitude is
boosted by powers of $k^2/(a H)^2=a (k/H_0)^2$ compared to the
naive expectation of powers of $\Phi$. This is in contrast to
backreaction in inflationary backgrounds
\cite{Tsamis, Mukhanov, Nambu:2000a, Nambu:2000b, Li:2001, Abramo, Nambu:2002, Geshnizjani:2002, Brandenberger:2002, Finelli:2003, Geshnizjani:2003},
%[40, 41, 45-53],
where gradients are negligible and the effect is boosted by
the large phase space of infrared terms. However, in both cases,
the backreaction can be large though the magnitude of
each individual mode $\Phi_{\bk}$ is small.

For linear perturbations, backreaction
of order $(k/a H)^2\Phi^2$ was found to have a magnitude 
of at most $10^{-5}$ today. Backreaction of order $(k/a H)^4\Phi^2$
reduces to a boundary term which is zero when evaluated over
the whole space with periodic boundary conditions. The magnitude
over the horizon volume with realistic boundary conditions is
unknown, but could be of order one, at least before structure
formation.
At second order in $\Phi$ there are no higher orders in momentum.
However, at higher orders there could be terms such as $(k/a H)^8\Phi^4$
that would not reduce to boundary terms and that would
straightforwardly contribute corrections of order one.

That perturbations might be important today is perhaps not
surprising. A measure of the inhomogeneity and anisotropy of spacetime
is given by the Weyl tensor. For the metric \re{metric},
the ratio of the square of the Weyl tensor to the square
of the scalar curvature is (for $a\propto t^{2/3}$)
\bea
  \frac{C_{\alpha\beta\gamma\delta} C^{\alpha\beta\gamma\delta}}{R^2} \simeq 
 \frac{8}{9}\frac{1}{(a H)^4} \left( \pat_i\pat_j\Phi\pat_i\pat_j\Phi - \frac{1}{3}\nabla^2\Phi\nabla^2\Phi \right) \ ,
\eea

\noindent which, when averaged, is essentially the integral
\re{back2}, in other words $\sigma^2$. That this ratio is
not small suggests that the impact of inhomogeneities can be large.

A few words about the backreaction framework are in order.
If the backreaction is sizeable, then we have
clearly exceeded the range of validity of the approximation
of taking the background and the perturbations from first
order perturbation theory. What would change in a consistent
second order calculation?

First of all, first order scalar perturbations will
in general give rise to second order vector and tensor
perturbations \cite{Finelli:2003, Acquaviva:2002, Matarrese:1997}, so 
the metric could not be written in the simple form \re{metric}.
Second, we should check higher order terms of the type
$(\pat_i\Phi\pat_i\Phi)^2$ and
$(\nabla^2\Phi\nabla^2\Phi)^2$ to verify that the
truncation to second order is consistent\footnote{Note that
the last three terms in \re{hubble} are strictly speaking
quartic in $\Phi$.}.

Ignoring these complications and assuming that
$\theta$ and $\tau$ would still be given by
\re{u}, \re{theta} and \re{tau}, the main change would be to
couple the development of the perturbations to their effect
on the background expansion rate. If the backreaction
led to faster expansion, this would be expected
to cause the perturbations in the linear
regime to decay, which would in turn slow down the
expansion (as well as affect the apparent equation of state).
As in the `concordance model', this would alleviate
the problem of too high $\sigma_8$ in models with $\Omega_{\textrm{m}}=1$,
so that neutrinos might not be needed to damp the power spectrum.
The backreaction would also be expected to have an impact on
the low multipoles of the CMB. Naively, one would expect
the decay of the gravitational potential to increase the
amplitude of the low multipoles just as in the `concordance
model'. However, without considering the second order calculation
in detail, it is not clear what the effect on the low
multipoles would actually be.

As noted earlier, a consistent second order calculation would probably
also answer the question of whether the perturbations are Gaussian with
respect to the physical perturbed spacetime or the background spacetime.

It is interesting that the backreaction naturally gives the
magnitudes $(k_{\textrm{eq}}/H_0)^2\Phi^2 H_0\sim 10^{-5} H_0$ and $H_0$.
Even the first one, though too small to explain the apparent acceleration,
is closer to $H_0$ than what would be expected from particle
physics motivated models of dark energy. Indeed, it seems more natural
for the scale of present-day acceleration to emerge from cosmology
and astrophysics rather than particle physics. In the present
perturbative backreaction framework, the impact of the perturbations
which are breaking away from the general expansion seems promising.
These could have a large effect on the expansion rate, with a negative
apparent equation of state, at the right time to solve
the coincidence problem.

\ack

I thank George Efstathiou, Pedro Ferreira, Fabio Finelli,
Enrique Gazta\~{n}aga, Subir Sarkar, Misao Sasaki
and James Taylor for helpful discussions, and Ruth Durrer
for comments on the manuscript.

I also thank the Department of Applied Mathematics and Theoretical
Physics at the University of Cambridge for their hospitality at
the workshop ``Cosmological Perturbations on the Brane'', during
which some of this work was done.

The research has been supported by PPARC grant PPA/G/O/2002/00479,
by a grant from the Magnus Ehrnrooth Foundation and by the
European Union network HPRN-CT-2000-00152,
``Supersymmetry and the Early Universe''.\\

\appendix

\setcounter{section}{1}

\end{document}